\RequirePackage{lineno}
\documentclass[twocolumn,nofootinbib,floatfix]{revtex4-1}

\usepackage{amsmath,amssymb}
\usepackage{cancel}
\usepackage{graphicx}
\usepackage{epsfig}
\usepackage{amsmath,amsfonts,amssymb}
\usepackage{colortbl}
\usepackage{hhline}
\usepackage{pifont}
\usepackage{multirow}

\begin{document}

\title{Pulling the Higgs and Top needles from the jet stack \\[1mm] with Feature Extended Supervised Tagging}

\author{J.~A.~Aguilar-Saavedra}
\affiliation{Departamento de F\'{\i}sica Te\'{o}rica y del Cosmos, Universidad de Granada, E-18071 Granada, Spain}

\begin{abstract}
Jet tagging has become an essential tool for new physics searches at the high-energy frontier. For jets that contain energetic charged leptons we introduce Feature Extended Supervised Tagging (FEST) which, in addition to jet substructure, considers the features of the charged lepton within the jet. With this method we build dedicated taggers to discriminate among boosted $H \to \ell \nu q \bar q$, $t \to \ell \nu b$, and QCD jets (with $\ell$ an electron or muon). The taggers have an impressive performance, allowing for overall light jet rejection factors of $10^4-10^5$, for top quark / Higgs boson efficiencies of $0.5$. The taggers are also excellent in the discrimination of Higgs bosons from top quarks and vice versa, for example rejecting top quarks by factors of $100-300$ for Higgs boson efficiencies of $0.5$. We demonstrate the potential of these taggers to improve the sensitivity to new physics by using as example a search for a new $Z'$ boson decaying into $Z H$, in the fully-hadronic final state.
\end{abstract}

\maketitle

\section{Introduction}
\label{sec:1}

From the last decade the Large Hadron Collider (LHC) is probing the high-energy frontier of particle interactions. With the high luminosity achieved, it has been possible to explore the multi-TeV scales not only in the search for new resonances, but also to test the SM production mechanisms at high energy, looking for possible deviations from the predictions of the Standard Model (SM). Being the two most massive SM particles, the Higgs boson and the top quark play a unique role in the search for physics beyond the SM, in particular to probe the electroweak symmetry breaking. The Higgs boson mainly decays hadronically or semileptonically (fully leptonic and diphoton decay modes are rare) while the top quark always produces a $b$ quark in its decay. Therefore, when they are produced with a large boost, their decay products merge into a single jet $J$.

Jet tagging has witnessed a tremendous progress in the last decade~\cite{Butterworth:2008iy,Thaler:2010tr,Larkoski:2014gra,Moult:2016cvt,Datta:2017rhs} (see Ref.~\cite{Larkoski:2017jix} for a review). The goal of the different tagging methods is to distinguish a `signal' jet resulting from the hadronic decay of a boosted heavy particle, such as a weak $W/Z$ boson, a Higgs boson, or a top quark, from a `background' quark or gluon jet. The discrimination is done by the analysis of the jet substructure: while the former jets are multi-pronged (containing two or three quarks, depending on the decaying particle) the latter only have one prong. Jet tagging methods have been extensively used, for instance, in searches for new gauge bosons, scalar and spin-2 particles~\cite{Sirunyan:2017ukk,Sirunyan:2017hsb,Aaboud:2018juj,Aaboud:2018zba,Aaboud:2018fgi,Sirunyan:2018ikr,Sirunyan:2019vxa,Sirunyan:2019jbg,Aaboud:2018eoy,Aad:2020ddw}, vector-like quarks~\cite{Sirunyan:2017ynj,Sirunyan:2018ncp,Aaboud:2018zpr,Sirunyan:2019xeh} and dark matter~\cite{{Sirunyan:2018gdw}}, as well as in SM measurements~\cite{Sirunyan:2017dgc,Sirunyan:2020hwz}. 

Generic supervised taggers have also been developed~\cite{Aguilar-Saavedra:2017rzt,Aguilar-Saavedra:2020sxp,Aguilar-Saavedra:2020uhm} aiming to distinguish arbitrary multi-pronged jets from QCD jets. They have been found capable of separating jets containing `prompt' (produced in the hard process) non-isolated leptons from QCD jets in which the leptons result from the decay of $b$, $c$ quarks. However, to the best of our knowledge, no tagger has been specifically developed for jets containing such leptons. (Notice, however, that non-isolated leptons are routinely used as one of the ingredients for $b$-tagging of jets.) 
This paper aims to fill that gap.  We build up on the previously introduced Mass Unspecific Supervised Tagging (MUST)~\cite{Aguilar-Saavedra:2020uhm} to develop neural network (NN) taggers which, in addition to jet mass, transverse momentum ($p_T$) and substructure variables, use as input the charged lepton energy fraction $z = E_\ell / E_J$ and the distance from the jet axis in the plane of pseudorapidity ($\eta$) and azimuthal angle $(\phi)$, $\Delta R = (\Delta \eta_{\ell J}^2 + \Delta \phi_{\ell J}^2)^{1/2}$. The method hereby introduced is dubbed as Feature Extended Supervised Tagging (FEST). We build dedicated taggers that can discriminate among $H \to \ell \nu q \bar q$, $t \to \ell \nu b$ and QCD jets, treating the $\ell = e,\mu$ cases separately. These two examples have the highest interest, since there are numerous measurements and searches by the ATLAS and CMS experiments involving top quarks or Higgs bosons in the boosted regime. We note that early work~\cite{Thaler:2008ju,Rehermann:2010vq} pointed out the usefulness of non-isolated leptons for the identification of $t \to \mu \nu b$. The related lepton $p_T$ fraction $\hat z = p_{T \ell} / p_{T J}$ has been shown~\cite{Aguilar-Saavedra:2014iga,Aguilar-Saavedra:2017vka,Aguilar-Saavedra:2019ptp} very useful to discriminate boosted top quarks from QCD jets. A variant, using the lepton $p_T$ fraction with respect to a sub-jet, was explored in Ref.~\cite{Brust:2014gia}, where a detailed study on lepton isolation was also performed. The electron energy fraction has also been indirectly used in Ref.~\cite{Chatterjee:2019brg}.


\section{Generating the event samples}
\label{sec:2}

 The Monte Carlo samples used to train and test the NNs are obtained as follows. Boosted Higgs bosons are generated with {\scshape MadGraph}~\cite{Alwall:2014hca}, in the SM process $pp \to ZH$, with $Z \to \nu \bar \nu$ and $H \to \ell \nu q \bar q$. For boosted top (anti-)quarks we use $pp \to Zt + Z\bar t$ mediated by a vector flavour-changing $tcZ$ coupling~\cite{delAguila:1999ac}, with $Z \to \nu \bar \nu$ and $t \to \ell \nu b$. 
For these processes the top flavour-changing neutral interactions are implemented in {\scshape Feynrules}~\cite{Alloul:2013bka} and interfaced to {\scshape MadGraph5} using the universal Feynrules output~\cite{Degrande:2011ua}. QCD jets are generated in the inclusive process $pp \to jj$, with $j$ a light jet (not including $b$ quarks). A possible extension could include $b \bar b$ in the training too; however, the tagger trained on light jets has excellent performance  for $b$ jets, as it is explicitly seen in the example presented in Section~\ref{sec:6}.

Event samples are generated in 100 GeV bins of $p_T$ starting at $[300,400]$ GeV, and up to to $[2.1,2.2]$ TeV in the case of QCD samples. For Higgs bosons and top quarks the jet $p_T$ is actually smaller than the $p_T$ of the decaying heavy particle, due to the missing neutrino. Therefore, we extend the generation up to the $[2.9,3.0]$ TeV and 
$[3.4,3.5]$ TeV bins, respectively. This guarantees coverage of the entire jet $p_T$ range up to 2.2 TeV. Even though within each bin the events mainly populate the lower end of the interval, the bins are narrow enough to adequately parameterise the $p_T$ dependence. For testing purposes, $b \bar b$ samples are generated using the same $p_T$ binning.

The parton-level event samples so generated are passed through {\scshape Pythia}~\cite{Sjostrand:2007gs} for hadronisation and {\scshape Delphes}~\cite{deFavereau:2013fsa} for a fast detector simulation, using the CMS card. Jets are reconstructed with {\scshape FastJet}~\cite{Cacciari:2011ma} applying the anti-$k_T$ algorithm~\cite{Cacciari:2008gp} with radius $R=0.8$, and groomed with Recursive Soft Drop~\cite{Dreyer:2018tjj}. In the subsequent analysis we only keep jets with groomed mass $m_J \in [40,170]$ GeV and $p_T \geq 400$ GeV. The chosen mass range encompasses the jet mass distributions for top quark and Higgs boson jets, and the latter cut is imposed in order to have a sufficient boost for top quarks, so that its decay products are contained within a $R=0.8$ jet. We also ask that the jets contain a charged lepton with $p_T \geq$ 10 GeV within a distance $\Delta R=0.8$ of the jet axis.
As discussed in the Appendix, the overall selection efficiencies for jet preselection plus tagging are quite independent of this mild lower cut. For top and Higgs high-$p_T$ jets, the leptons are already very energetic and the lepton $p_T$ threshold has little influence. On the other hand, for QCD jets a higher threshold at preselection significantly lowers the efficiency. However, the NNs eventually learn that leptons are much softer for QCD jets, and a lower preselection efficiency is compensated by a higher mistag rate by the tagger.

We note that the requirement to contain a lepton, even with a threshold as low as $p_T \geq 10$ GeV, has a very low efficiency for the QCD jet samples. With our simulation we find that, for example, for the sample with $p_T \in [1,1.1]$ TeV at the partonic level the efficiencies to find an electron or a muon above this threshold are 0.041 and 0.020, respectively. Therefore, huge samples of dijet events are needed to have sufficient statistics: $4 \times 10^5$ events per $p_T$ bin for NN training and validation, and $6 \times 10^5$ for testing, totaling 19 million $jj$ pairs.


\section{Building the taggers}
\label{sec:3}

 Jet substructure is characterised by the set of $N$-subjettiness variables proposed in~\cite{Datta:2017rhs},
 \begin{equation}
 \left\{ \tau_1^{(1/2)}, \tau_1^{(1)}, \tau_1^{(2)}, \dots , \tau_{5}^{(1/2)}, \tau_{5}^{(1)}, \tau_{5}^{(2)}, \tau_{6}^{(1)}, \tau_{6}^{(2)} \right\} \,,
 \label{ec:taulist}
 \end{equation}
computed for the ungroomed jets.\footnote{We note that the performance might be improved by using low-level jet substructure variables. For top quarks decaying hadronically, it has been shown~\cite{Kasieczka:2019dbj} that taggers using low-level variables achieve a background rejection $\sim 1.4$ times larger than taggers using $N$-subjettiness.}
By means of a principal component analysis, it can be seen that the number of physically relevant combinations is actually smaller. Still, because the computational speed is not a serious issue, we keep the above set. As done in Ref.~\cite{Aguilar-Saavedra:2020uhm}, we include as NN inputs the jet mass, but varying on a narrower range $m_J \in [40,170]$ GeV, and the jet $p_T \in [0.4,2.2]$ TeV. Moreover, as previously pointed out, for these taggers we also include the lepton energy fraction $z$ and $\Delta R$ with respect to the jet. A standardisation of the 21 inputs, based on the SM background distributions, is performed to improve the NN learning.

Our goal is to simultaneously discriminate among jets corresponding to Higgs bosons, top quarks and light quarks / gluons. Therefore, we build NNs whose input are the aforementioned variables for jets corresponding to the three classes ($H$, $t$, $j$).  The NN output is a list of three numbers $(p_1,p_2,p_0)$, with $p_1 + p_2 + p_0 = 1$, giving the probabilities that a jet corresponds to the $H$, $t$ or $j$ class, respectively. The NNs contain two hidden layers of 512 and 64 nodes, with Rectified Linear Unit (ReLU) activation for the hidden layers and a softmax function for the outputs. The NNs are optimised with the categorical cross-entropy loss function, using the Adam~\cite{Kingma:2014vow} optimiser. Two independent NNs are built, for $\ell = e$ and $\ell = \mu$, using {\scshape Keras}~\cite{keras} with a {\scshape TensorFlow} backend~\cite{tensorflow}.
The training sets for the $e$ ($\mu$) NN contain 6000 (5000) events from each class ($H$, $t$, $j$) and $p_T$ slice, totaling around $3 \times 10^5$ training events. The validation sets used to monitor the NN performance have similar size and composition as the training ones.

For testing, we build additional two-class NNs to discriminate between (i) $H$ and $j$; (ii) $t$ and $j$; (iii) $H$ and $t$, using the same architecture except for the loss function, for which we use the (binary) cross-entropy, and the output layer, which only contains one node with a sigmoid activation function. These NNs are trained only using the events corresponding to the two classes ($H$, $j$), ($t$, $j$) or ($H$, $t$), respectively. Furthermore, we also build NNs only using the jet mass and $p_T$, and the charged lepton $z$ and $\Delta R$ as input, to investigate to which extent the jet substructure variables contribute to the discrimination. 

Let us finally mention here some checks concerning the NN architecture. We have not found any performance improvement when duplicating the size of the first hidden layer. In previous work~\cite{Aguilar-Saavedra:2020uhm} we also verified that including higher-order $\tau_n^{(\beta)}$ does not improve the tagger discrimination. We also investigated the possibility of using unbalanced samples in the training, or other generalised loss functions such as the one proposed in~\cite{Murphy:2019utt}, without noticeable improvements.


\section{Tagger performance}
\label{sec:4}

\begin{figure}[t]
\begin{center}
\begin{tabular}{c}
\includegraphics[width=8cm,clip=]{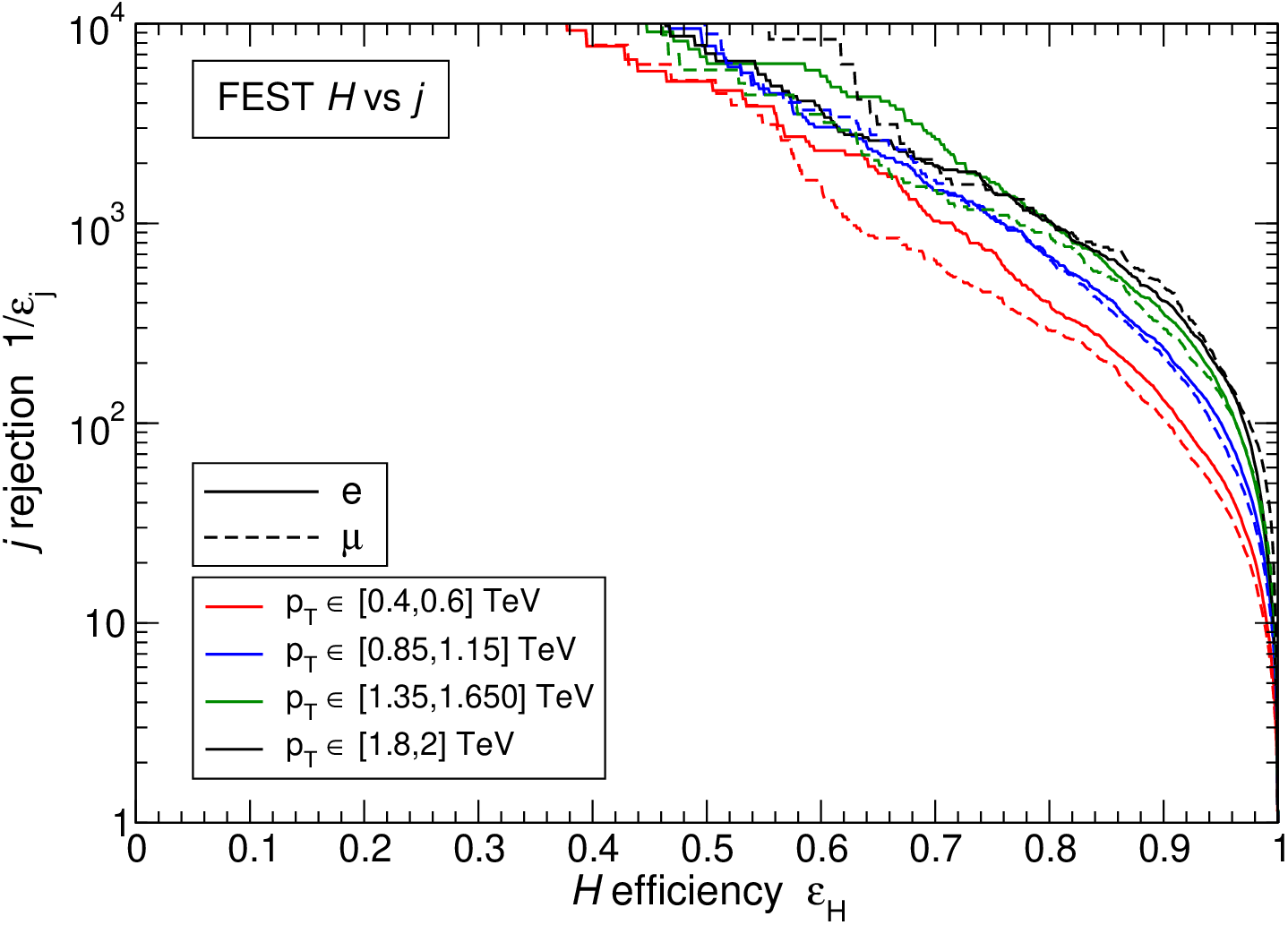} \\[2mm]
\includegraphics[width=8cm,clip=]{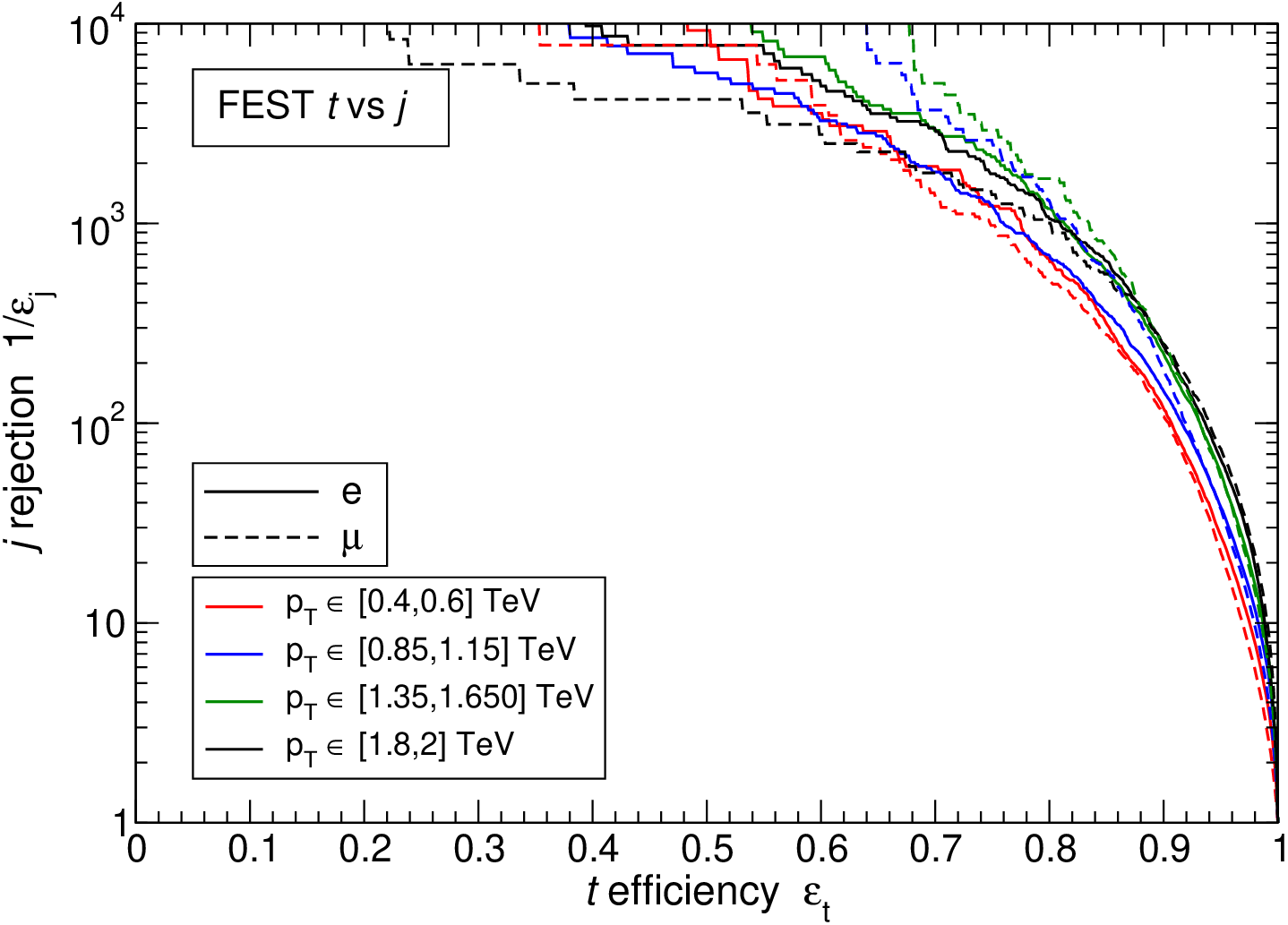}  \\[2mm]
\includegraphics[width=8cm,clip=]{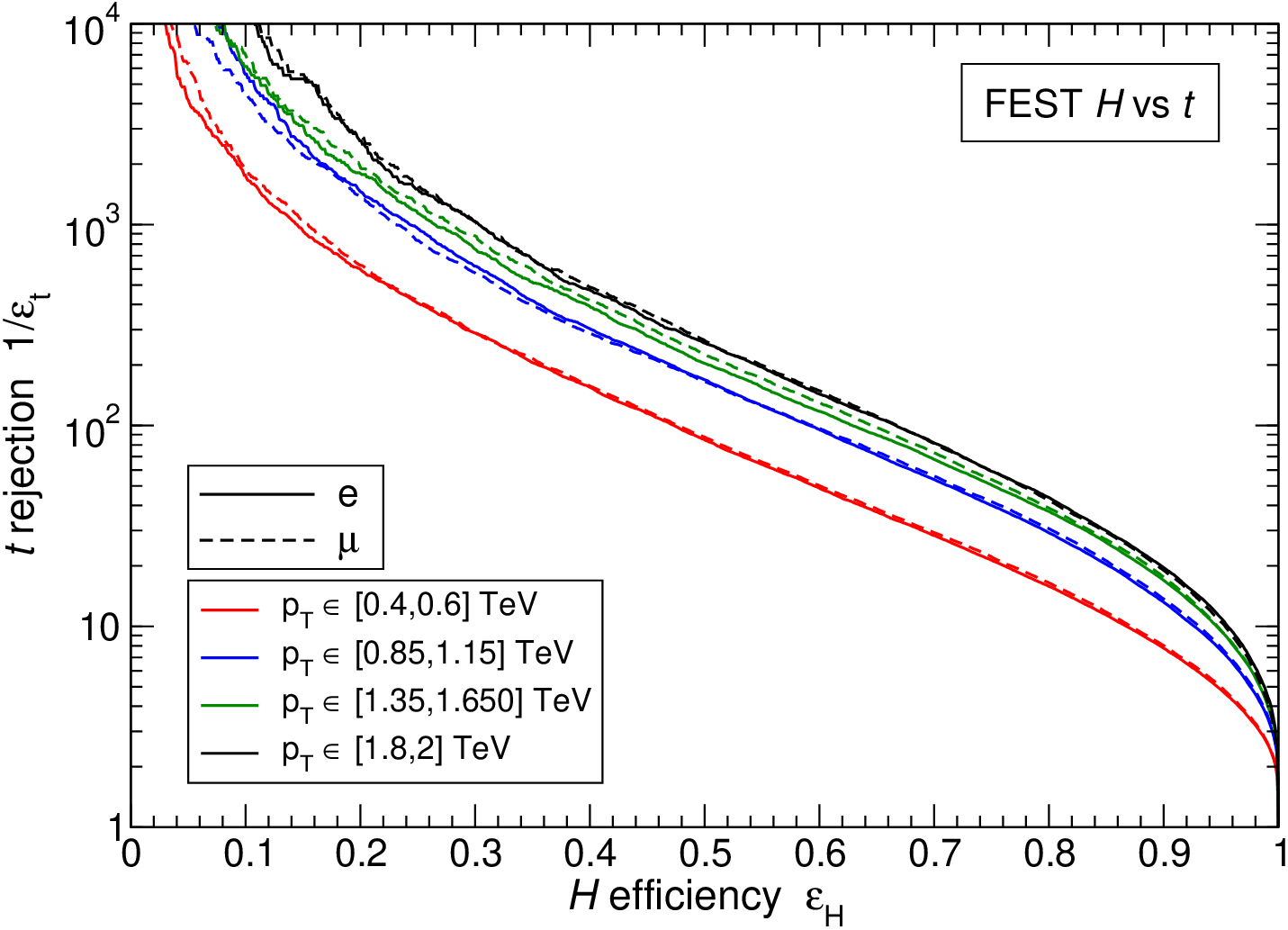}
\end{tabular}
\caption{ROC curves for the separation between different jet classes (for details see the legends and the main text).}
\label{fig:ROC}
\end{center}
\end{figure}

 We test the ability of our taggers to discriminate between different pairs of classes, marginalising over the third one. Figure~\ref{fig:ROC} shows the receiver operating characteristic (ROC) curves for $H$ versus $j$ (top), $t$ versus $j$ (middle) and $H$ versus $t$ (bottom). In all plots, the horizontal axis gives the tagging efficiency $\varepsilon$ for a given type of jet, and the vertical axis the tagging rejection $\varepsilon^{-1}$ for another type of jet. In $H$ versus $t$  we consider $t$ as `background' because Higgs boson production is not usually a background for top quark measurements, but the discrimination can be performed in either way. The ROCs are shown for jets in four $p_T$ intervals: $[0.4,0.6]$, $[0.85,1.15]$, $[1.35,1.65]$ and $[1.8,2]$ TeV.\footnote{The test samples have a few tens of thousands of events, therefore for rejection factors above $5 \times 10^3$ the statistical fluctuations may become important, especially at high transverse momentum.} The area under the ROC curve (AUC) is very high in all cases, reaching values around 0.998 for $H$ versus $j$, 0.995 for $t$ versus $j$ and 0.98 for $H$ versus $t$, for transverse momenta around 2 TeV.
 
\begin{figure}[t]
\begin{center}
\begin{tabular}{c}
\includegraphics[width=8cm,clip=]{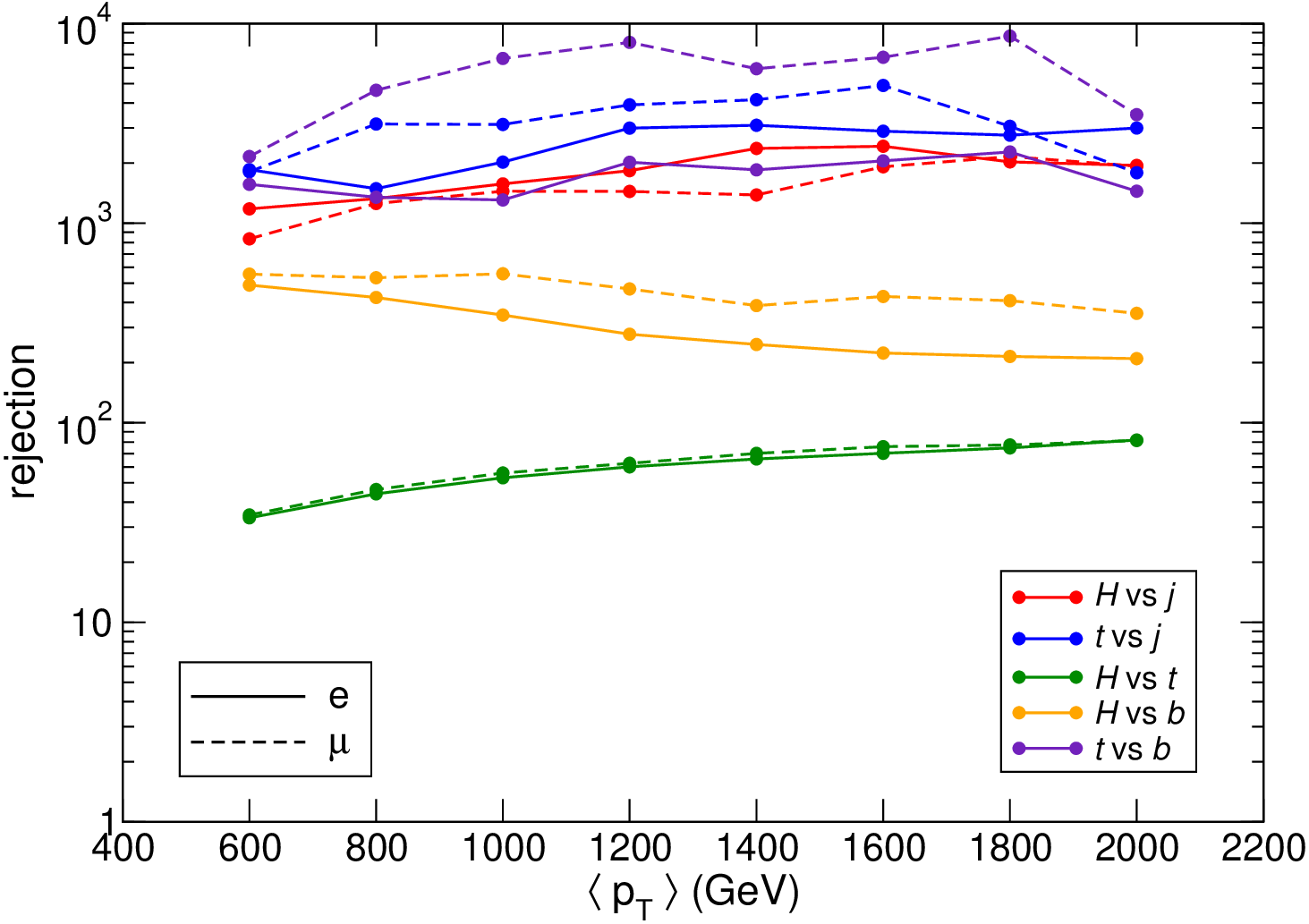} 
\end{tabular}
\caption{Rejection factors for $H$ versus $j$ (red), $t$ versus $j$ (blue) and $H$ versus $t$ (green) as a function of transverse momentum for fixed tagging efficiencies of 0.7 (see the text for details).}
\label{fig:REJ}
\end{center}
\end{figure}

Figure~\ref{fig:REJ} shows the rejection factors $\varepsilon^{-1}$ for fixed efficiencies of 0.7, as a function of the jet $p_T$. The efficiencies are evaluated within intervals of $p_T \in [ \langle p_T \rangle - 200, \langle p_T \rangle + 200]$ GeV and plotted as a function of $\langle p_T \rangle$. We also include here lines corresponding to the discrimination against $b$-quark jets, which have not been used in the NN training. As it can be readily seen, the discrimination of both $H$ and $t$ jets from $b$ jets is excellent, and likely sufficient to reject backgrounds involving $b$ quarks.

The tagger rejection for QCD jets is impressive. Furthermore, let us remind the reader that the test samples, for which the ROC curves in figure~\ref{fig:ROC} and rejection factors in figure~\ref{fig:REJ} are computed, are composed of jets that already pass the preselection requirement of a charged lepton with $p_T \geq 10$ GeV.  And for QCD jets, the efficiency of this lepton requirement is quite small (see the Appendix). For a given overall $H$ efficiency $\bar \varepsilon_H$, the overall QCD jet rejection $\bar \varepsilon_j^{-1}$ is straightforwardly calculated as follows:\footnote{We use a bar to distinguish the overall efficiencies, including preselection, from the tagger efficiencies $\varepsilon$. The overall efficiency for $H$ and $t$ jets is defined relative to the full $H \to \ell \nu q \bar q$ and $t \to \ell \nu b$ samples (within some $p_T$ range) before preselection, not summing over lepton flavours. Likewise, the overall efficiency for QCD jets is computed relative to the full sample within some $p_T$ range.}
\begin{itemize}
\item By dividing the selected overall efficiency $\bar \varepsilon_H$ by the preselection efficiency (either for electrons or for muons) we get a $H$ tagging efficiency $\varepsilon_H$, to which corresponds a $j$ rejection $\varepsilon_j^{-1}$.
\item Then, dividing $\varepsilon_j^{-1}$ by the preselection efficiency for QCD jets (either for electrons or muons), we obtain the overall QCD jet rejection factor $\bar \varepsilon_j^{-1}$.
\end{itemize}
For example, the preselection efficiencies for $H \to \ell \nu q \bar q$ jets with $p_T \in [1,1.1]$ TeV are 0.61 and 0.91 in the electron and muon channel, respectively. For QCD jets, they are 0.041 and 0.020. Therefore, considering jets with $p_T \sim 1$ TeV, for an overall efficiency $\bar \varepsilon_H = 0.5$,
the corresponding light jet rejection factors are
$$
\begin{array}{cccccc}
e: & \varepsilon_H = 0.61 & \rightarrow & \varepsilon_j^{-1} = 3000  & \rightarrow & \bar \varepsilon_j^{-1} = 7.4 \times 10^4 \\[2mm]
\mu: & \varepsilon_H = 0.55 & \rightarrow & \varepsilon_j^{-1} = 4400  & \rightarrow & \bar \varepsilon_j^{-1} = 2.2 \times 10^5 
\end{array}
$$
These overall rejection factors of the order of $10^5$ for QCD jets make the tagger quite useful, even if the decays $H \to \ell \nu q \bar q$ are subdominant.

Similar comments can be made regarding the top jet discrimination from QCD jets. The preselection efficiencies for $t$ jets with $p_T \in [1,1.1]$ TeV are 0.73 and 0.80 in the electron and muon channel, respectively. Therefore, for an overall $t$ efficiency $\bar \varepsilon_t = 0.5$, the corresponding light jet rejection factors $\bar \varepsilon_j^{-1}$ are
$$
\begin{array}{cccccc}
e: & \varepsilon_t = 0.68 & \rightarrow & \varepsilon_j^{-1} = 2000  & \rightarrow & \bar \varepsilon_j^{-1} = 4.8 \times 10^4 \\[2mm]
\mu: & \varepsilon_t = 0.62 & \rightarrow & \varepsilon_j^{-1} = 11000  & \rightarrow & \bar \varepsilon_j^{-1} = 5.5 \times 10^5 
\end{array}
$$
As expected, the QCD jet rejection is much larger than for the top fully-hadronic decay. For reference, NN taggers for the hadronic top quark decay mode have a light jet rejection factor of 500 for a top tagging efficiency of 0.5, working in the same $p_T$ range~\cite{Kasieczka:2017nvn,Macaluso:2018tck}. (Note that neither of these taggers, nor the FEST tagger presented here, use $b$ tagging to identify top quarks.) Of course, the figures are not comparable because they refer to different decay modes. A meaningful comparison can be made considering the improvement on the $S/\sqrt B$ ratio (with $S$ standing for signal and $B$ for background) brought by the different taggers, also taking into account the branching ratio for the hadronic and leptonic modes,
\begin{align}
& t \to q \bar q b:  && \text{Br}(t \to q \bar q b) \frac{\varepsilon_t}{\sqrt \varepsilon_j} = 7.5 \,, \notag \\
& t \to e \nu b: && \text{Br}(t \to e  \nu b) \frac{\varepsilon_t}{\sqrt \varepsilon_j} = 12 \,, \notag \\
& t \to \mu \nu b: && \text{B}r(t \to \mu  \nu b) \frac{\varepsilon_t}{\sqrt \varepsilon_j} = 40 \,.
\label{ec:fom}
\end{align}
With this figure of merit, one can see that tagging the top semileptonic decays with FEST offers much better prospects to probe for new physics.

The discrimination between $H$ and $t$ jets is also excellent, as seen in the lower panel of Fig.~\ref{fig:ROC}, and this is of high importance because top quark production may constitute a background to Higgs boson measurements, as will be seen in the $Z' \to ZH$ example presented in the following.


\section{Comparison with two-class taggers}
\label{sec:5}

\begin{figure}[t]
\begin{center}
\begin{tabular}{c}
\includegraphics[width=8cm,clip=]{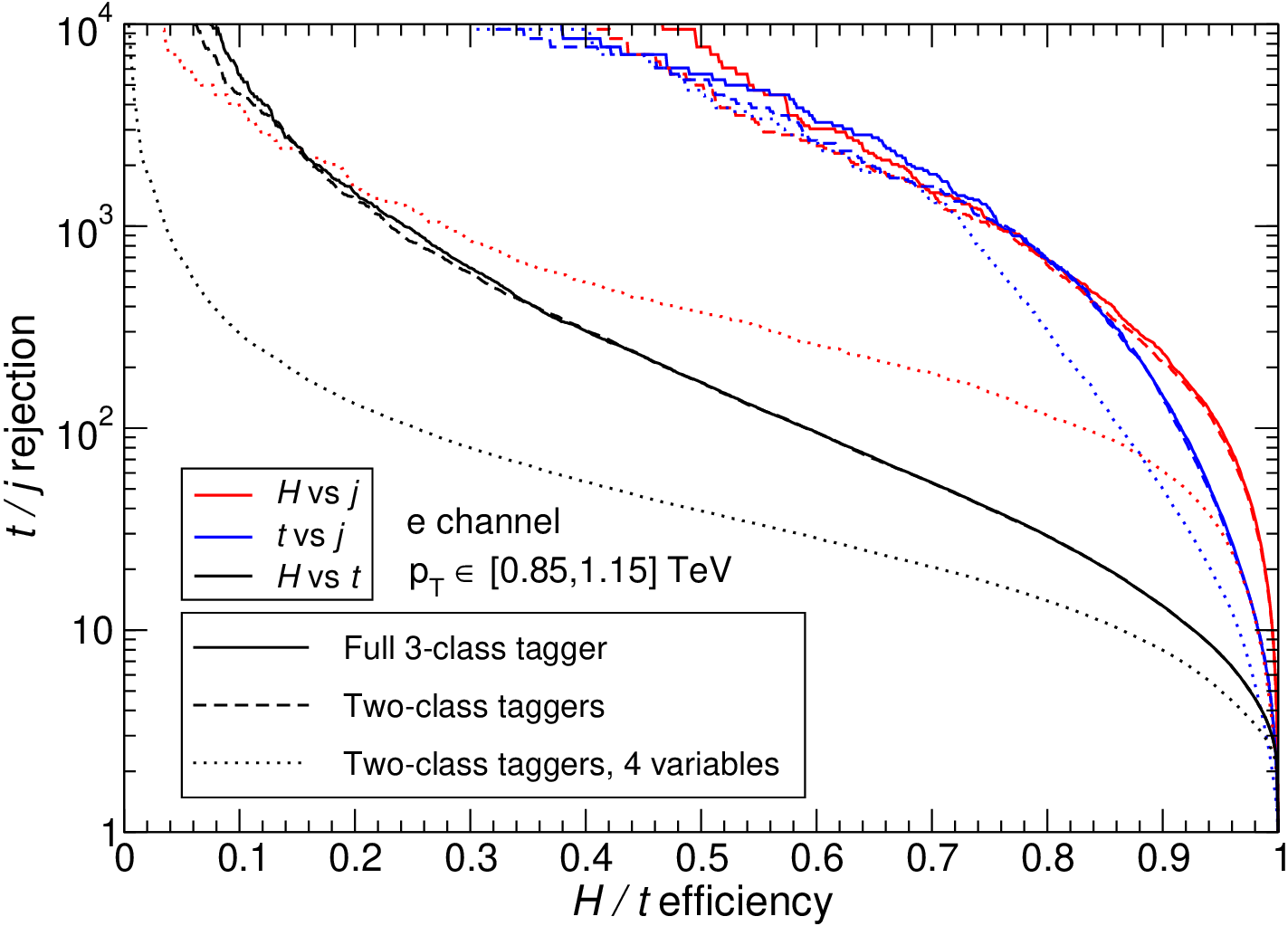} 
\end{tabular}
\caption{Comparison of ROC curves for different taggers (for details see the legends).}
\label{fig:comp}
\end{center}
\end{figure}

We restrict ourselves to the electron channel and the test interval $p_T \in [0.85,1.15]$ TeV to compare the three-class tagger discriminating among $H$, $t$ and $j$, with less general two-class taggers. The results are shown in Fig.~\ref{fig:comp}. Interestingly, the discrimination power is the same for the three-class and the two-class taggers, with minor differences that may well have a statistical nature. This fact shows that the discrimination power between two given classes is not degraded when building a tagger that simultaneously tries to distinguish among $H$, $t$ and $j$.

Because the lepton energy (or $p_T$) fraction has previously been used as a simple discriminating variable between top quarks decaying semileptonically and QCD jets \cite{Thaler:2008ju,Rehermann:2010vq,Aguilar-Saavedra:2014iga,Aguilar-Saavedra:2017vka,Aguilar-Saavedra:2019ptp,Brust:2014gia,Chatterjee:2019brg}, it is worth exploring to which extent the jet substructure variables add to the discrimination. With this purpose, we build two-class taggers that only use as input the jet mass and $p_T$, as well as $z$ and $\Delta R$. As expected, for $H \to \ell \nu qq$ (with two quarks) the jet substructure significantly enhances the discrimination with respect to light jets. For $t \to \ell \nu b$, jet substructure variables help but are less important. For $H$ versus $t$ discrimination the analysis of the jet substructure is crucial, as expected, because the former jets have two quarks and the latter only one. 

Conversely, as seen in Refs.~\cite{Aguilar-Saavedra:2020sxp,Aguilar-Saavedra:2020uhm}, generic taggers only using substructure variables have a poorer discrimination between jets with leptons and QCD jets. The tests in those references are performed using as signal jets from boosted heavy neutrinos decaying $N \to e q \bar q$, but the conclusion is expected to be general. 

\section{Example: $Z' \to ZH$}
\label{sec:6}

\begin{figure}[htb]
\begin{center}
\begin{tabular}{c}
\includegraphics[width=8cm,clip=]{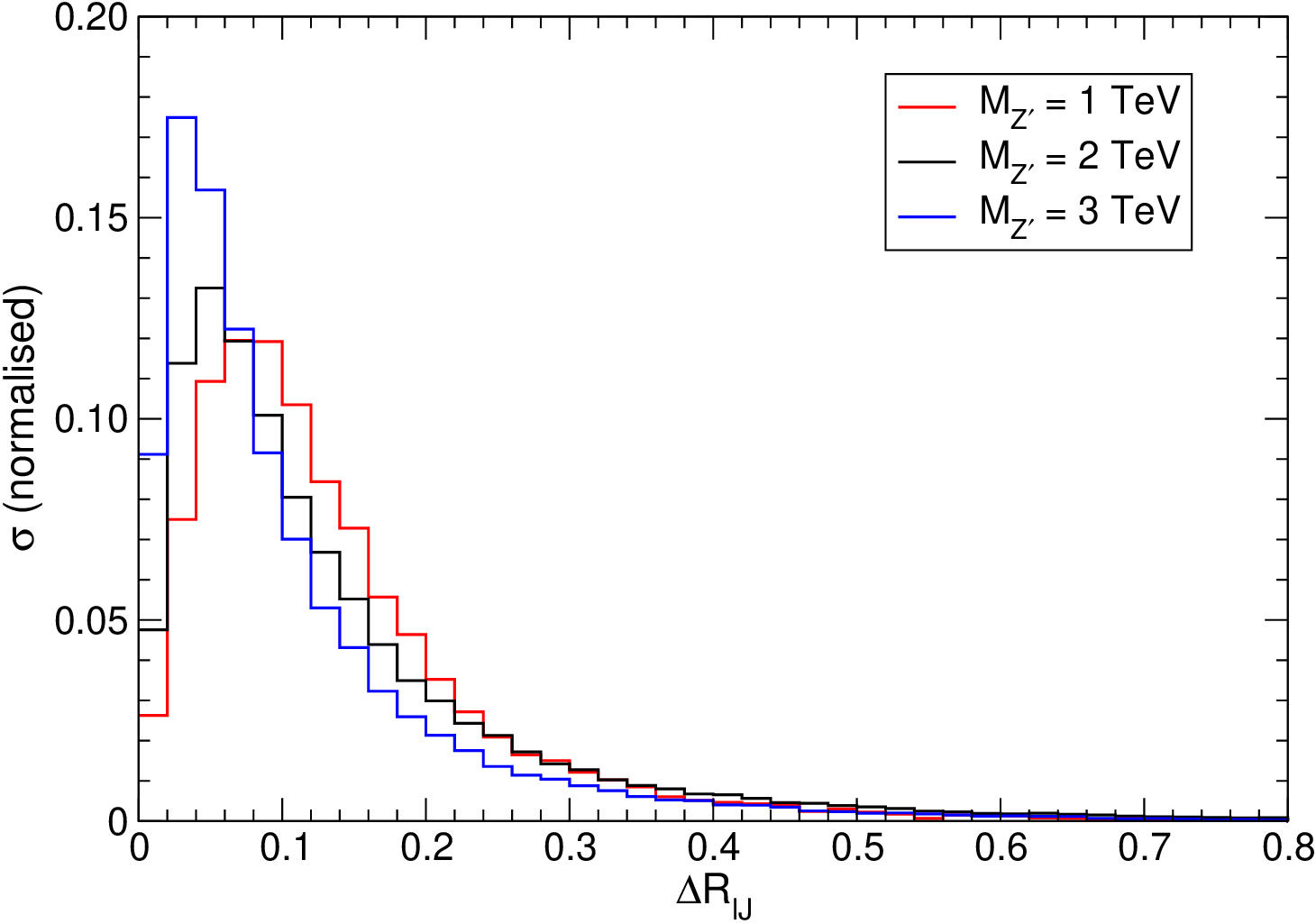} 
\end{tabular}
\caption{$\Delta R$ separation between the charged lepton and the Higgs jet produced in the $Z' \to ZH$ decay.}
\label{fig:dR}
\end{center}
\end{figure}

We investigate here the usefulness of the taggers here introduced to improve the sensitivity of LHC measurements.  
Tagging of boosted $H \to b \bar b$ is performed both by the ATLAS and CMS Collaborations by looking at $b$-tagged subjets of a large-radius jet containing  the $H \to b \bar b$ decay products. Namely, the ATLAS Collaboration uses $R=0.2$ subjets in earlier searches~\cite{Aaboud:2017ahz} and variable radius jets in the most recent one~\cite{Aad:2020tps} with the full Run 2 dataset. The CMS Collaboration uses subjets of $R=0.4$~\cite{Sirunyan:2021bzu}. Requiring one or two $b$-tagged subjets significantly suppresses the QCD background, especially in the latter case. The ATLAS Collaboration has considered the decay $H \to \ell \nu q \bar q$ in a search for $HH$ resonances~\cite{ATLAS:2018fpd} in the resolved case, where this decay produces two narrow $R = 0.4$ jets and a charged lepton that can be independently reconstructed. As the Higgs bosons are more boosted, the efficiency of the resolved final state decreases and the final state where all $H$ decay products are merged into a single jet becomes more sensitive. This can be seen in Fig.~\ref{fig:dR}, where we show the $\Delta R$ separation between the charged lepton and the axis of the jet containing the $H$ decay products in $Z' \to ZH$, $H \to \ell \nu q \bar q$. We select three different $Z'$ masses to illustrate the dependence on the heavy resonance mass. Because the lepton isolation criterion requires the absence of significant energy in a cone of radius $\Delta R \sim 0.1$ around the charged lepton, the resolved channel is disfavoured for resonances beyond the TeV scale. Future studies are required to compare the sensitivity of the resolved and merged final states for boosted $H \to \ell \nu q \bar q$.
 
Our goal here is to evaluate the potential sensitivity of new physics searches targeting the $H \to \ell \nu q \bar q$ decay in the merged final state, tagged using FEST. The branching ratio $\text{Br}(H \to \ell \nu q \bar q) = 0.13$ (summing over $\ell = e,\mu$ and lepton charges) is much smaller than $\text{Br}(H \to b \bar b) = 0.58~$\cite{deFlorian:2016spz} but the excellent performance of the FEST tagger makes the decay mode competitive for large luminosities, and especially in final states where the background is large. Otherwise, the large background rejection achieved by FEST is less useful.

We investigate the sensitivity of $ZH$ resonance searches in the decay modes $Z \to q \bar q$, $H \to \ell \nu q \bar q$. This fully-hadronic final state also allows to show the usefulness of the tagger to simultaneously suppress backgrounds with light jets and top quarks --- at the end the latter turn out to be the dominant ones. We  take as our reference for comparison the search for $ZH$ resonances in the fully-hadronic channel by the ATLAS Collaboration with the full Run 2 dataset~\cite{Aad:2020tps}, focusing on the $Z \to q \bar q$, $H \to b \bar b$ decay modes. Because our results are obtained with fast simulation, the comparison with the sensitivity achieved in Ref.~\cite{Aad:2020tps} has the caveat of a possible degradation of the tagger performance in the environment of a real experiment, therefore the comparison has to be taken with a grain of salt.

We perform a simulation including the backgrounds from $jj$, $t \bar t$, $Wjj$ and $tW$ production. Potential backgrounds with fake leptons cannot be handled with the fast simulation, but we expect them not to be dominant. In any case, in an experimental analysis they must be included. The dijet sample is the same one used to test the NN performance, and $t \bar t$, $Wjj$ and $tW$ samples are also generated in the same 100 GeV slices of $p_T$. Samples with $p_T \geq 2.2$ TeV are also generated, and the different samples are combined with weight proportional to the cross section. A 2 TeV $Z' \to ZH$ signal is generated with $Z \to q \bar q$, $H \to \ell \nu q \bar q$. For $M_{Z'} = 2$ TeV, the 95\% confidence level upper limit on the production cross section times decay branching ratio from Ref.~\cite{Aad:2020tps} is $\sigma(pp \to Z' \to ZH) \leq 5.3$ fb. We use this cross section as reference for comparison between the two $H$ decay channels.
Events are passed through the simulation chain described before. In addition to $R=0.8$ jets, we use a collection of `track jets' of radius $R = 0.2$, reconstructed using only tracks. A jet is considered as $b$-tagged if a $b$-tagged track jet (using the 70\% efficiency working point) within the $R=0.8$ jet is found. 

\begin{figure}[t!]
\begin{center}
\begin{tabular}{c}
\includegraphics[width=8cm,clip=]{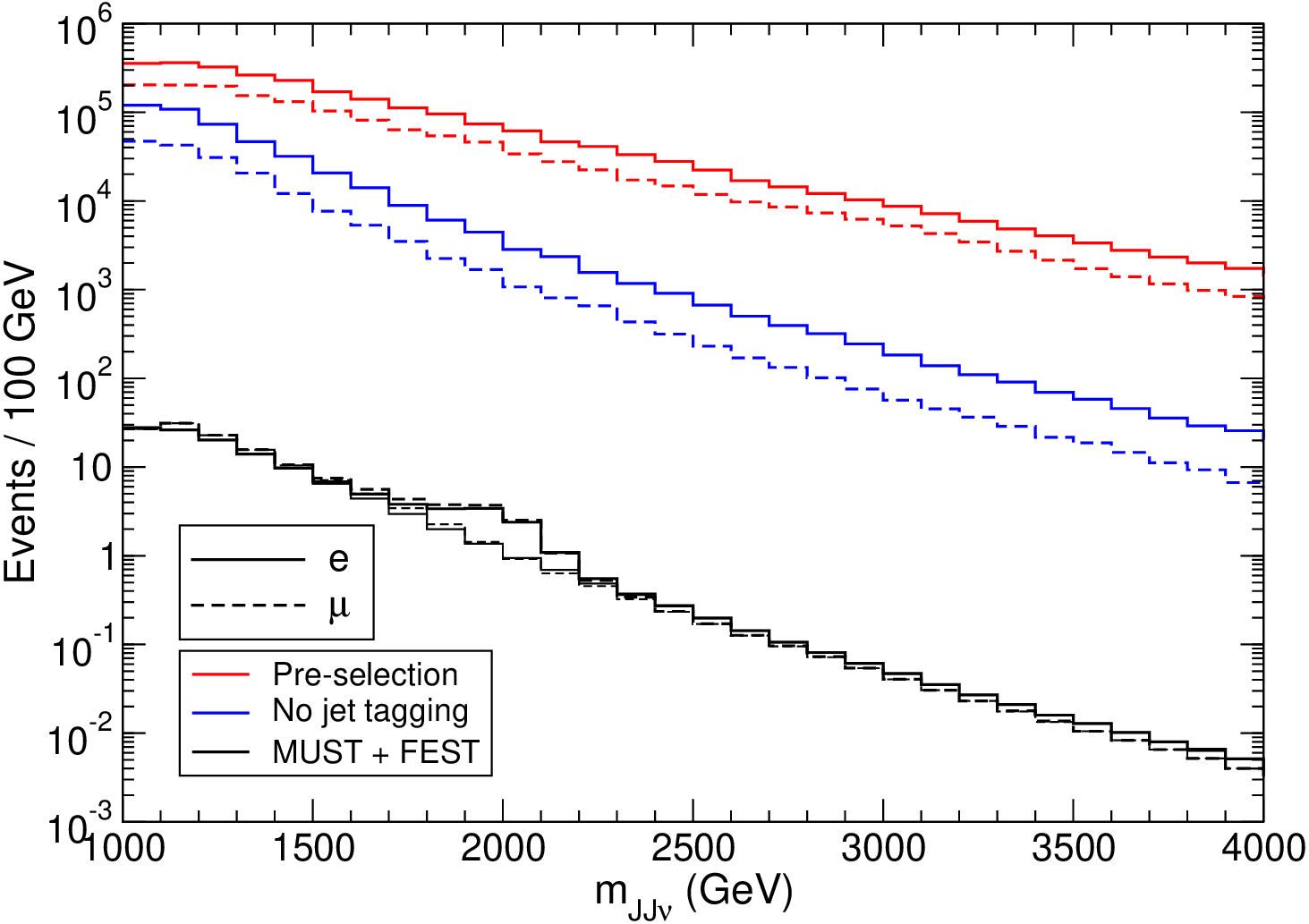} 
\end{tabular}
\caption{Reconstructed $Z'$ mass at different levels of event selection, for the background plus an injected signal.}
\label{fig:mJJn}
\end{center}
\end{figure}

As event preselection we require two jets with $m_J \geq 40$ GeV, $p_T \geq 400$ GeV and $|\eta| \leq 2.5$. At least one of them is required to have a charged lepton inside the jet. That jet is labeled as the `$H$' jet; if both jets have charged leptons, the one having the lepton with highest $z$ is selected. The remaining jet is labeled as `$Z$'. As a proxy for the $Z'$ mass we use the invariant mass of the two jets plus the neutrino, $m_{JJ\nu}$. The neutrino three-momentum is taken parallel to the one of the charged lepton, with its transverse component equal to the missing energy in the event.\footnote{We have also explored an alternative neutrino momentum reconstruction,  with the longitudinal component and energy determined by requiring that the invariant mass of the neutrino and the $H$ jet equal the Higgs boson mass. This constraint yields a second degree equation; among the two solutions we choose the one that gives smaller longitudinal momentum. The results with this alternative reconstruction are slightly worse.}
The $m_{JJ\nu}$ distribution for the background (overwhelmingly $jj$) at preselection is shown in Fig.~\ref{fig:mJJn}, normalised to a luminosity of 139 fb$^{-1}$.

Before jet tagging, we require a separation $|\Delta \eta| \leq 1.5$ among the two jets, jet masses $m_J \leq 110$ GeV, and perform a $b$-tag veto on the $H$ jet. These simple cuts reduce the background (which still is dominated by $jj$ production) by a factor of $10-100$, as shown in Fig.~\ref{fig:mJJn}.

Finally, tagging of both jets is performed. For the $H$ jet we require probabilities $p_0 \leq 0.01$, $p_2 \leq 0.9$ that the jet corresponds to the $j$ and $t$ class, respectively. For the $Z$ jet we use the two-pronged MUST-based tagger {\tt T$_\text{2P}$} developed in Ref.~\cite{Aguilar-Saavedra:2020uhm}, requiring a NN score (quantifying the probability that the jet is two-pronged) $X \geq 0.8$. Tagging the $H$ jet reduces the dijet background by a factor of $2.8 \times 10^{-3}$, and tagging the $Z$ jet reduces it by an additional factor of $0.04$. Thus, the tagging reduces the background by $3-4$ orders of magnitude, as shown in figure~\ref{fig:mJJn}, and allows the injected $Z'$ signal to be seen as a bump in the falling $m_{JJ\nu}$ distribution. For clarity, the background-only distributions after tagging are shown as thin lines. 
\begin{table}[h]
\begin{center}
\begin{tabular}{ccc}
  & $e$ & $\mu$ \\
$Z'$ & 3.5 & 3.9 \\
$t \bar t$ & 0.75 & 1.12 \\
$Wjj$ & 0.87 & 0.87 \\
$jj$ & 0.51 & 0.17 \\
$tW$ & 0.18 & 0.19 
\end{tabular}
\caption{Expected number of events for the signal and backgrounds in the bins with $m_{JJ\nu} \in [1.9,2.1]$ TeV, for a luminosity of 139 fb$^{-1}$.}
\label{tab:bkg}
\end{center}
\end{table}
After tagging, the expected number of events for the signal and the different backgrounds near 2 TeV is given in Table~\ref{tab:bkg}. Other backgrounds from $Zj$ and $Wj$ production, with $Z/W$ hadronic decay, are less important, and $b \bar b$ is even smaller. At the region near 2 TeV, the former two amount to $1/7$ and $1/3$ of the $jj$ background in the electron and muon channel, respectively, and the latter to $1/20$ and $1/9$, with the final event selection. 

The expected significance of the $Z'$ signal can be computed by performing likelihood tests for the presence of narrow resonances over the expected background, using the CLs method~\cite{Read:2002hq} with the asymptotic approximation of Ref.~\cite{Cowan:2010js}. The local significance at $m_{JJ\nu} = 1.95$ TeV is of $2.2\sigma$ in the $e$ channel and $2.4\sigma$ in the $\mu$ channel, neglecting systematic uncertainties.\footnote{Because the background after event selection at the signal region amounts to a handful of events, we expect background systematic uncertainties to be much smaller than the statistical uncertainty itself. On the other hand, for the signal it is in principle possible to calibrate the tagging efficiency in samples involving boosted Higgs bosons.} Combining both, the local significance reaches $3.2\sigma$. Therefore, even having in mind that the comparison with full simulation is not fair, it seems likely that the sensitivity to $Z' \to ZH$ may be improved, or at least matched, by the $H \to \ell \nu q \bar q$ decay mode.


\section{Concluding remarks}
\label{sec:7}

We have developed a three-class tagger to discriminate among boosted $H \to \ell \nu q \bar q$, $t \to \ell \nu b$, and light jets, with an impressive rejection rate for the latter, and excellent discrimination between top quarks and Higgs bosons. For top quarks, its possible applications are numerous, because the huge rejection factor for light jets overly compensates the smaller semileptonic decay branching ratio. Using as figure of merit the branching ratio times significance improvement, c.f.~(\ref{ec:fom}), tagging top quarks in the electron and muon channels improves over the hadronic decay mode previously considered by factors of 1.6 and 5, respectively. For Higgs boson the prospects are quite good too, despite the smaller branching fraction for $H \to \ell \nu q \bar q$.

Our tagger has been built to work on a very wide range of jet $p_T \in [0.4,2.2]$ TeV. (In contrast, several hadronic top taggers in the literature~\cite{Kasieczka:2017nvn,Macaluso:2018tck} are trained with jets within a narrow $p_T$ range.) This interval is sufficiently large so as to demonstrate that the tagger can correctly learn to distinguish the differences in jet substructure arising from different $p_T$ regimes and from different jet prongness. Moreover, it has been shown in Ref.~\cite{Aguilar-Saavedra:2020uhm} that the performance of a tagger trained on wide intervals of jet mass and $p_T$ nearly matches the performance of a tagger trained on narrow intervals. Therefore, the arbitrarily chosen range $p_T \in [0.4,2.2]$ TeV can be further extended and we do not expect a performance drop. 

One possible caveat to the practical application of the tagger is the possible difficulty and uncertainties in the measurement of $z$ and $\Delta R$ for electrons embedded within jets, and the possible appearance of fakes. Reference~\cite{Brust:2014gia} performed a detailed study regarding electron isolation, and there are good prospects that the measurements will be feasible. But even in a worst-case scenario that measurements in the electron channel could not be performed --- which, we stress again, seems unlikely --- the sensitivity in the muon channel alone is better than in hadronic top decays, c.f.~(\ref{ec:fom}), and likewise is expected for Higgs decays, as shown in the previous section.

Generally, one expects that $H \to \ell \nu q \bar q$ and $t \to \ell \nu b$ with the taggers here introduced will provide the best sensitivity for boosted Higgs boson and top quark measurements, except at the kinematical end of the spectrum where the background is already quite small. Therefore, for large integrated luminosities, and especially at the high-luminosity upgrade of the LHC, tagging these decay modes may provide the best sensitivity for boosted $H$, $t$ measurements across a very wide kinematical range.

\begin{figure}[t!]
\begin{center}
\begin{tabular}{c}
\includegraphics[width=8cm,clip=]{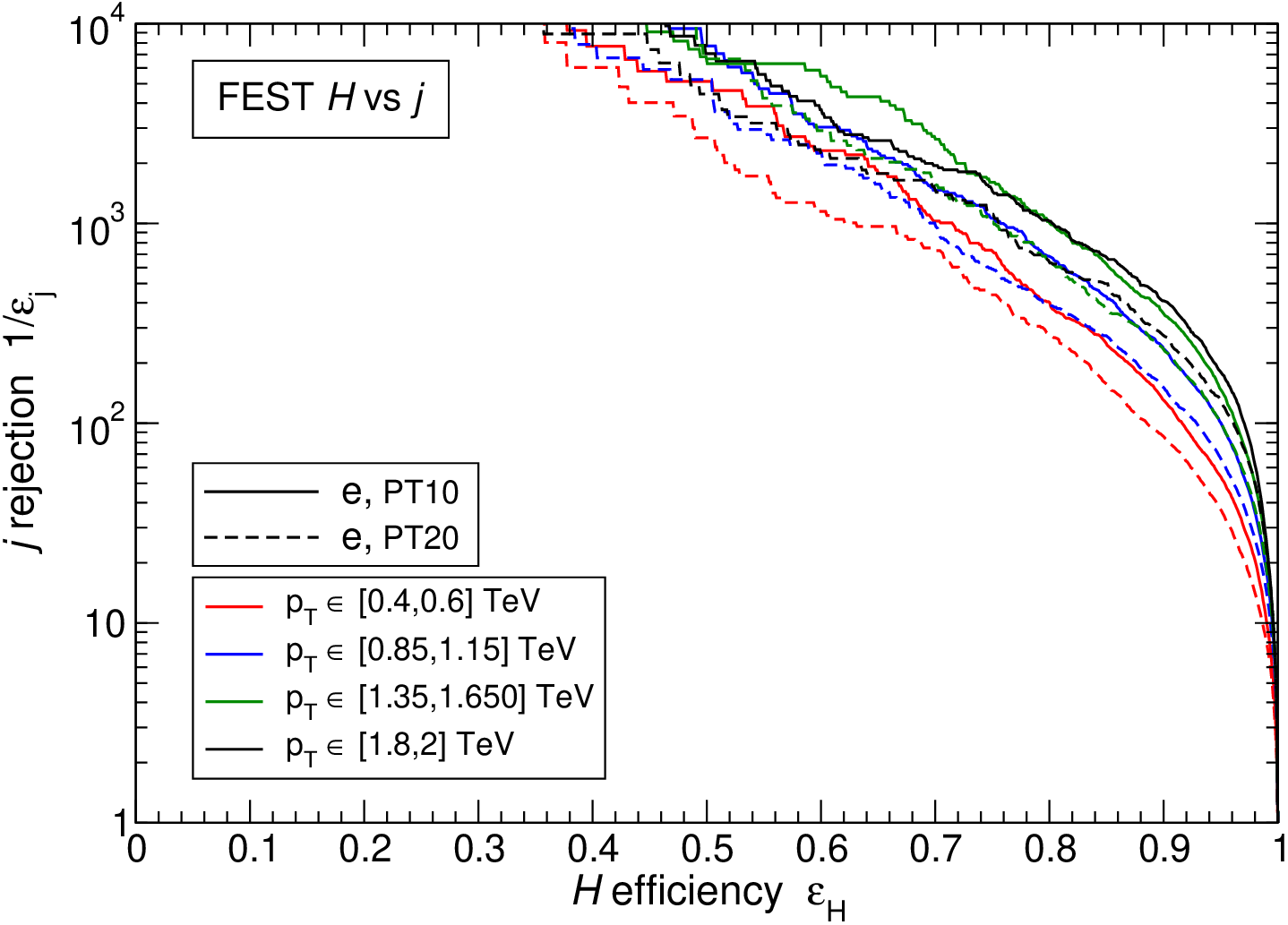} \\[2mm]
\includegraphics[width=8cm,clip=]{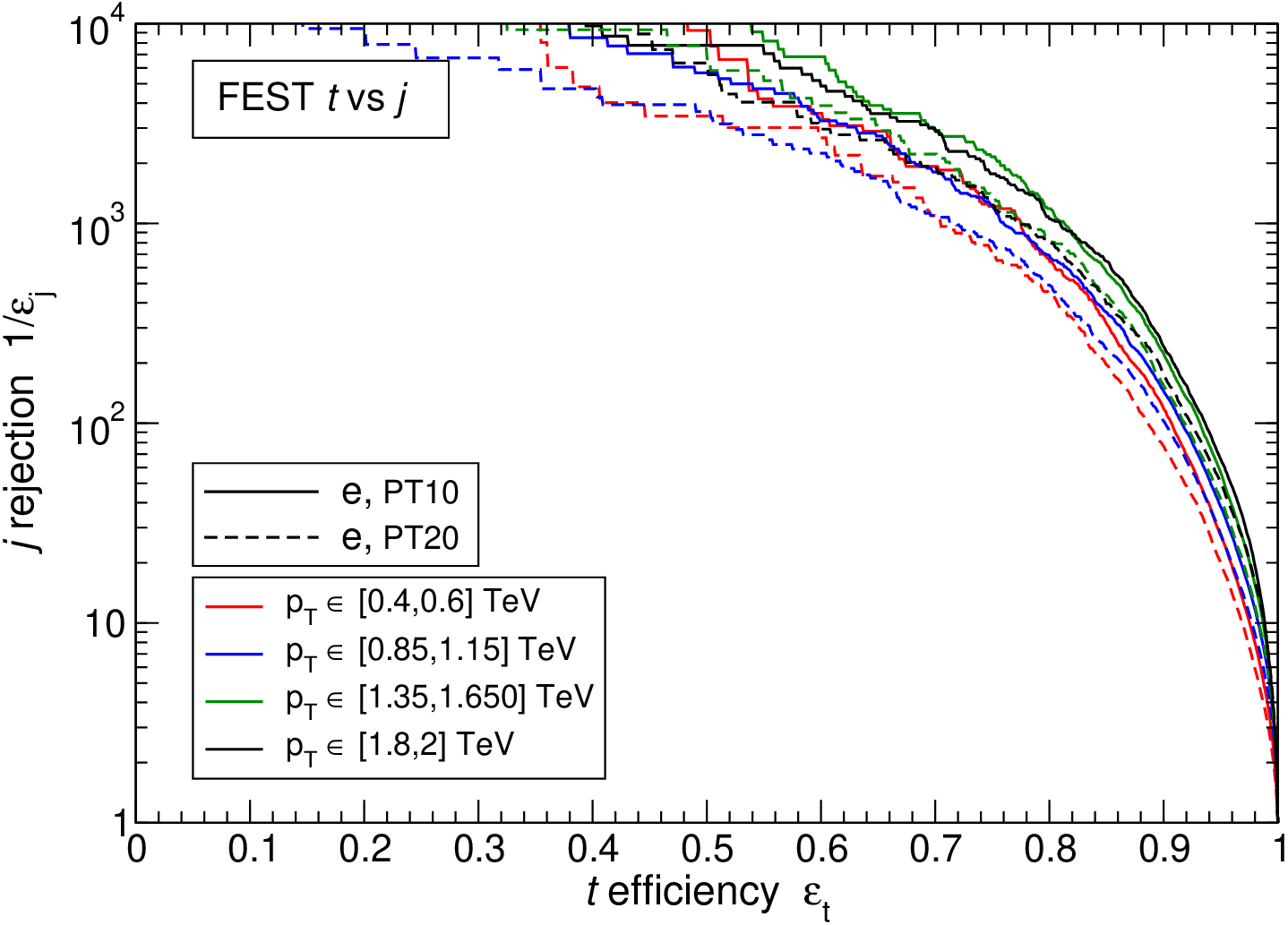}  \\[2mm]
\includegraphics[width=8cm,clip=]{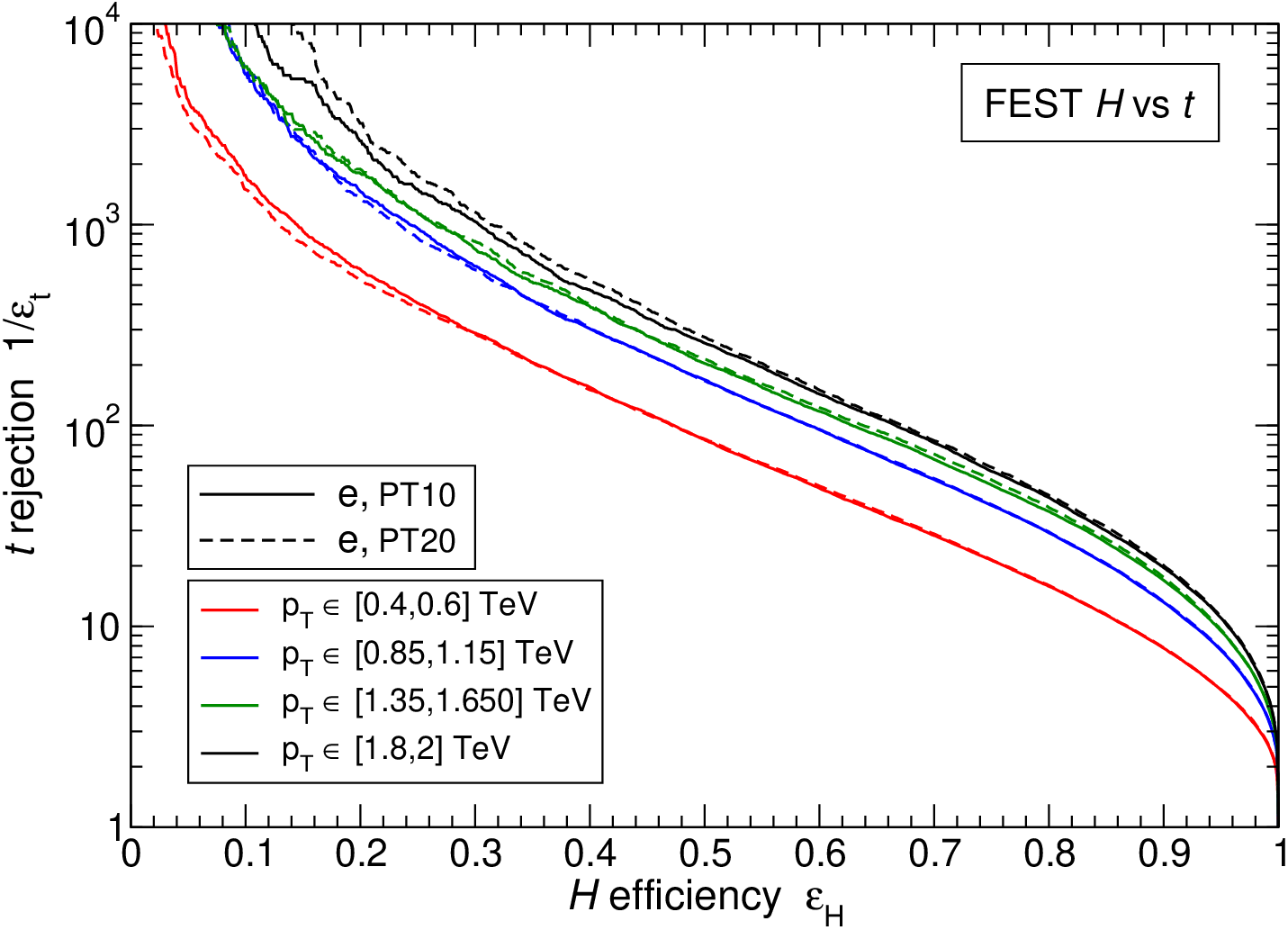}
\end{tabular}
\caption{Comparison of the ROC curves for the separation between different jet classes, for the samples with event preselection $p_T \geq 10$ GeV (as used throughout the paper) and $p_T \geq 20$ GeV.}
\label{fig:ROC-comp}
\end{center}
\end{figure}

Finally, let us comment that more generic taggers for jets containing leptons can also be built, which could be sensitive for example to boosted heavy neutrinos decaying $N \to \ell q \bar q$, and may be presented elsewhere.

\section*{Acknowledgements}

I thank J. Aguilar Saavedra for the use of computing resources, and F.R. Joaquim and J. Seabra for previous colaboration in the MUST development. This work has been supported by MICINN project PID2019-110058GB-C21 and by FCT project CERN/FIS-PAR/0004/2019.

\appendix
\section{Overall performance and event preselection}
\label{sec:a}

The tagger is built based on a sample of jets that already contain a charged lepton, with a minimum transverse momentum $p_T \geq 10$ GeV. As it has been argued, the overall performance should have little dependence on this choice, within reasonable limits. In this appendix we explicitly test this, by restricting ourselves to the electron channel and using jet samples that contain electrons with $p_T \geq 20$ GeV. The preselection efficiencies for jets of the three classes are collected in Table~\ref{tab:eff}. The same procedure is followed to train the NN, and the results are compared in Fig.~\ref{fig:ROC-comp} with the results previously obtained.
We denote by {\tt PT10} and {\tt PT20} the taggers built using electron thresholds $p_T \geq 10$ GeV, $p_T \geq 20$ GeV, respectively.
As expected, the performance in $H$ versus $j$ and $t$ versus $j$ jets in the ROC plots is degraded, since the higher preselection threshold already makes part of the work of the tagger in separating $H$ and $t$ (with energetic electrons) from $j$. Also as expected, the discrimination between $H$ and $t$ is practically unaltered, up to small differences arising from the use of different NNs. 

\begin{table}[htb]
\begin{center}
\begin{tabular}{cccc}
& $H$ & $t$ & $j$ \\
$p_T \geq 10$ GeV & 0.81 & 0.73 & 0.041 \\
$p_T \geq 20$ GeV & 0.80 & 0.72 & 0.023 \\
\end{tabular}
\caption{Preselection efficiencies for Higgs ($H$), top ($t$) and QCD ($j$) jets with $p_T \in [1,1.1]$ TeV, with the requirement to contain an electron with $p_T$ above the given threshold.}
\label{tab:eff}
\end{center}
\end{table}

Still, as argued in Section~\ref{sec:2}, the overall performance of the tagger is nearly independent of the lepton $p_T$ threshold.  Let us calculate for example the $j$ rejection for jets with $p_T \sim 1$ TeV, for an $H$ overall efficiency $\bar \varepsilon_H = 0.5$, as done in Section~\ref{sec:4}. For the two taggers, we have
$$
\begin{array}{cccccc}
\text{\tt PT10}: & \varepsilon_H = 0.61 & \rightarrow & \varepsilon_j^{-1} = 3000 & \rightarrow & \bar \varepsilon_j^{-1} = 7.4 \times 10^4 \\[2mm]
\text{\tt PT20}: & \varepsilon_H = 0.63 & \rightarrow & \varepsilon_j^{-1} = 1900 & \rightarrow & \bar \varepsilon_j^{-1} = 8.3 \times 10^4 
\end{array}
$$
The $O(10\%)$ difference in the overall light jet rejection factor is due to statistical fluctuations in the jet samples, caused by the high value of $\varepsilon_j^{-1}$. Likewise, can test the light jet rejection for an overall $t$ efficiency $\bar \varepsilon_t = 0.5$, 
$$
\begin{array}{cccccc}
\text{\tt PT10}: & \varepsilon_t = 0.68 & \rightarrow & \varepsilon_j^{-1} = 2000 & \rightarrow & \bar \varepsilon_j^{-1} = 4.8 \times 10^4 \\[2mm]
\text{\tt PT10}: & \varepsilon_t = 0.69 & \rightarrow & \varepsilon_j^{-1} = 1100 & \rightarrow & \bar \varepsilon_j^{-1} = 4.9 \times 10^4 
\end{array}
$$
and in this case the $j$ rejection is nearly the same when using either preselection threshold.

\end{document}